\documentclass[a4paper]{article}

\usepackage{INTERSPEECH2020}

\title{Multimodal Semi-supervised Learning Framework for Punctuation Prediction in Conversational Speech}
\name{Monica Sunkara, Srikanth Ronanki, Dhanush Bekal, Sravan Bodapati, Katrin Kirchhoff}
\address{Amazon AWS AI}
\email{\{sunkaral, ronanks\}@amazon.com}

\begin{document}

\maketitle
\begin{abstract}

In this work, we explore a multimodal semi-supervised learning approach for punctuation prediction by learning representations from large amounts of unlabelled audio and text data. Conventional approaches in speech processing typically use forced alignment to encoder per frame acoustic features to word level features and perform multimodal fusion of the resulting acoustic and lexical representations. As an alternative, we explore attention based multimodal fusion and compare its performance with forced alignment based fusion. Experiments conducted on the Fisher corpus show that our proposed approach achieves $\sim$6-9\% and $\sim$3-4\% absolute improvement (F1 score) over the baseline BLSTM model on reference transcripts and ASR outputs respectively. We further improve the model robustness to ASR errors by performing data augmentation with N-best lists which achieves up to an additional $\sim$2-6\% improvement on ASR outputs. We also demonstrate the effectiveness of semi-supervised learning approach by performing ablation study on various sizes of the corpus. When trained on 1 hour of speech and text data, the proposed model achieved $\sim$9-18\% absolute improvement over baseline model. 

 
\end{abstract}

\section{Introduction}

The output text generated from automatic speech recognition (ASR) systems is typically devoid of punctuation and sentence formatting. Lack of sentence segmentation and punctuation makes it difficult to comprehend the ASR output. For example, consider the two sentences: ``Let's eat Grandma'' vs. ``Let's eat, Grandma!''. Punctuation restoration not only helps understand the context of the text but also greatly improves the readability. Punctuated text often helps in boosting the performance of several downstream natural language understanding (NLU) tasks.

There is a plethora of work done in punctuation prediction over the past few decades. While some early methods of punctuation prediction used finite state or hidden markov models \cite{fst1,fst2}, some other techniques have investigated probabilistic models like language modeling \cite{lm2, lm3, lm5}, conditional random fields (CRFs) \cite{CRF2, CRF1} and maximum entropy models \cite{mem1}. As neural networks gained popularity, several approaches have been proposed based on sequence labeling and neural machine translation \cite{oktem2017attentional}. These models widely used convolutional neural networks (CNNs) and LSTM based architectures \cite{pahuja2017joint}. More recently, attention \cite{ottokar2015lstm, tilk2016bidirectional} and transformer \cite{yi2019self, nguyen2019fast, sunkara2020robust} based architectures which have been successfully applied to a wide variety of tasks, have shown to perform well for punctuation prediction.

Although it is a well explored problem in the literature, most of these improvements do not directly translate to all domains. In particular, punctuation prediction for conversational speech is not very well explored \cite{zelasko2018punctuation,augustyniak2020punctuation, sunkara2020robust}. Also, a number of approaches have been proposed exploiting the use of acoustic features in addition to lexical features for punctuation task, but they are rather limited and do not clearly address the gap in performance with ASR outputs. In this paper, we focus on multimodal semi-supervised deep learning approach for punctuation prediction in conversational speech by leveraging pretrained lexical and acoustic encoders.


\subsection{Relation to prior multimodal work}
\label{ssec:prior_work}

While several methodologies used either text or acoustic only information \cite{moro2017prosody} for predicting punctuation, many studies show that combining both the features yields the best performance \cite{yi2019self, klejch2016punctuated, christensen2001punctuation, steve2017sequenceprosody}. Acoustic features widely used in the literature include prosodic information such as pause duration, phone duration, and pitch related values like fundamental frequency, and energy. \cite{christensen2001punctuation} shows that using acoustic information lead to increased recognition of full stops. In \cite{steve2017sequenceprosody}, a hierarchical encoder is used to encode per frame acoustic features to word level features and the results show that incorporating acoustic features significantly outperform purely lexical systems. However, when trained on a very large independent text corpus, the lexical system outperformed the multimodal system that was trained on parallel audio/text corpora. To mitigate this, the work in \cite{yi2019self} introduced speech2vec embeddings but they do not vary with respect to the acoustic context in reference speech. 

In general, we identify two potential shortcomings with aforementioned multimodal systems. First, the training is still suboptimal due to lack of large-scale parallel audio/text corpora. Secondly, the models trained on reference text transcripts do not perform that well on ASR outputs, although incorporating acoustic features reduced the gap to some extent. 

\subsection{Novelty of this work}
\label{ssec:novelty}

In this work, we introduce a novel framework for multimodal fusion of lexical and acoustic embeddings for punctuation prediction in conversational speech. Specifically, we investigate the benefits of using lexical and acoustic encoders that are pretrained on large amounts of unpaired text and audio data using unsupervised learning.  The key idea is to learn contextual representations through unsupervised training where substantial amounts of unlabeled data is available and then improve the performance on a downstream task like punctuation, for which the amount of data is limited, by leveraging learned representations. For multimodal fusion, we explore attention mechanism to automatically learn the alignment of word level lexical features and frame level acoustic features in the absence of explicit forced alignments.

We also show the adaptation of our proposed multimodal architecture for streaming usecase by limiting the future context. We further study the effect of pretrained encoders with respect to varying data sizes and their performance when trained on very small amounts of data. Finally, we exploit the N-best lists from ASR to perform data augmentation and reduce the gap in performance when tested on ASR outputs. 




\section{Semi-supervised learning architecture}
\label{sec:ssl}

This section introduces our proposed \textit{multimodal semi-supervised learning architecture} (MuSe) for punctuation prediction. We pose the prediction task as a sequence labeling problem where the model outputs a sequence of punctuation labels given text and corresponding audio. The architecture contains three main components: acoustic encoder, lexical encoder, and a fusion block to combine outputs from both the encoders. Figure \ref{fig:ss_punc} shows a schematic overview of the proposed approach. 

The lexical encoder is pretrained on a large unlabelled text corpus for learning rich contextual representations and finetuned for the downstream task (i.e., weights are updated during punctuation model training). Given a sequence of input words ($x_1^l, x_2^l,..., x_m^l$), subwords ($s_1^l, s_2^l,..., s_n^l$) are extracted using a wordpiece tokenizer \cite{wordpiece}. The resulting subwords are fed as input to a pretrained encoder, which outputs a sequence of lexical features: $H^l$ = ($h_1^l, h_2^l,..., h_n^l$ ) at its final layer. 

The acoustic encoder takes audio signal as an input and outputs a sequence of frame level acoustic embeddings ($x_1^a, x_2^a,..., x_T^a$). The acoustic encoder is pretrained on a large unlabelled audio data with the objective to predict future samples from a given signal context. This unsupervised pretraining is based on the work of Schneider et al. \cite{schneider2019wav2vec}. After pretraining, we freeze the parameters of the acoustic encoder. The frame level acoustic embeddings are then passed through a convolution layer followed by a uni-directional LSTM layer to learn task specific embeddings: $\widetilde{H}^a$ = ($\Tilde{h}_1^a, \Tilde{h}_2^a,..., \Tilde{h}_T^a$). 

Since lexical and acoustic features differ in sequence length, it is not straightforward to concatenate them. Section \ref{sec:acoustic_fusion} discusses about two different approaches for aligning acoustic feature sequence with lexical sequence. Once we obtain the resulting aligned acoustic sequence $H^a$ = ($h_1^a, h_2^a,..., h_n^a$), we concatenate last layer representations of pretrained lexical encoder ($H^l$) with outputs from acoustic encoder ($H^a$) and input to a linear layer with softmax activation to classify over the punctuation labels generating ($\hat{p}_1, \hat{p}_2,..., \hat{p}_n$) as outputs.
\begin{equation}
    \hat{p_i} = softmax(W^k(h_i^l \oplus h_i^a) + b^k)
\end{equation}
\noindent where $W^k$, $b^k$ denote weights and bias of linear output layer. The model is finetuned end-to-end to minimize the cross-entropy loss between the predicted distribution ($\hat{p}_i$) and targets ($p_i$).

\vspace{-2mm}
\section{Multimodal fusion alignment}
\label{sec:acoustic_fusion}


Several approaches have been proposed in the past for fusion of acoustic features with lexical encoder. Most of these approaches used word-level prosodic inputs and concatenated with lexical inputs or outputs from lexical encoder. In this section, we describe how we model frame-level acoustic features for fusion with sub-word lexical encoder using two different approaches: using force-aligned word durations and sub-word attention model. 

\subsection{Forced alignment fusion}
\label{ssec:force_alignment}

Some prosodic features like fundamental frequency and energy can be averaged across each word and used as input to the acoustic encoder. Similarly, word duration can also be used as a feature and the work by \cite{zelasko2018punctuation} has shown minor improvements in punctuation prediction for conversational speech by employing relative word timing and duration of word. However, such mechanism does not capture the acoustic context beyond a word and also prevents the use of frame-level acoustic features where the average vector does not represent anything. 

For this reason, we model frame-level acoustic features using an LSTM-based acoustic encoder where force-aligned word durations are used to obtain final word boundaries. We then use word boundaries to select the respective LSTM state outputs to form word-level features. We duplicate the same output to all the sub-words within each word.

\begin{figure}[t]
\begin{minipage}[b]{1.0\linewidth}
  \centering
  \centerline{\includegraphics[width=8cm]{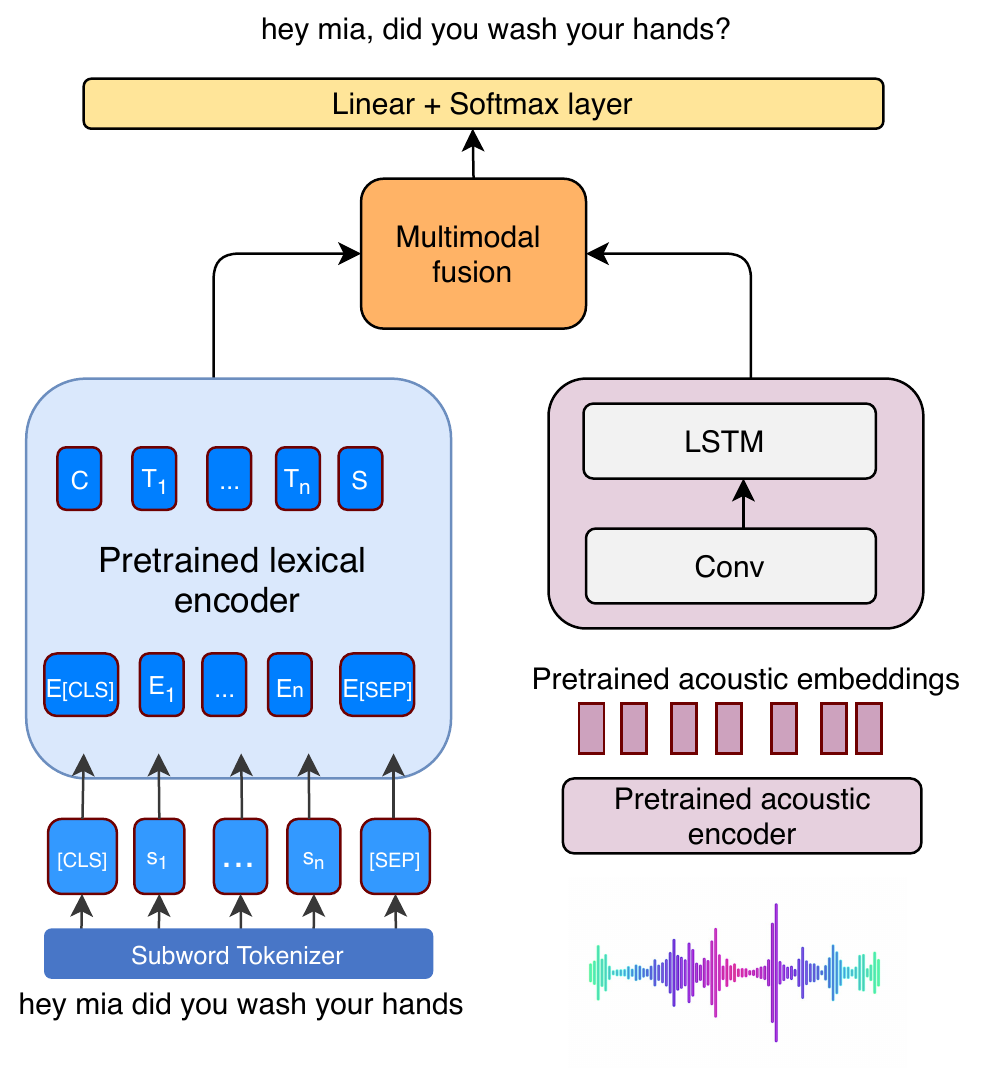}}
  \vspace{-4mm}
\caption{An overview of our  multimodal semi-supervised learning architecture for punctuation prediction.}
\vspace{-5mm}
\label{fig:ss_punc}
\end{minipage}
\end{figure}

\subsection{Attention fusion}
\label{ssec:attn_fusion}


In the previous approach, we require force-aligned durations during training and word-level timestamps at inference time to form sub-word acoustic embeddings. While this is possible to achieve with conventional hybrid ASR systems, it may become an overhead when used in conjunction with end-to-end ASR systems\footnote{Only in cases where punctuation is not modelled along with phonetic unit} such as LAS and Transformer \cite{chan2015listen, dong2018speech}. The use of force-aligned durations may limit the acoustic context to a limited number of frames. For this reason, we introduce an attention module that uses scaled dot-product attention \cite{vaswani2017attention} to find the alignment between acoustic feature sequence with sub-word lexical sequence, operating on query Q, key $\kappa$, and value $\vartheta$ :

\begin{equation}
    Attention(Q,\kappa,\vartheta) = softmax(\frac{Q\kappa^T}{\sqrt{d_\kappa}})\vartheta
\end{equation}

\noindent where $d_\kappa$ is the dimension of the keys. For attention model, we use the same encoder architecture as shown in Figure \ref{fig:ss_punc}. However, since attention may not require such a low-resolution input acoustic sequence, we downsample the input feature sequence by using a fixed stride of 2 in the 1-dimensional convolution layer. In the attention module, key $\kappa$ and value $\vartheta$ are obtained from LSTM state outputs. Sub-word encoder outputs are used as query for this attention. The key $\kappa$ is obtained by 
using a projection layer whose weight matrix is $W^\kappa$: 
\begin{equation}
    \kappa_i^a = f(W^\kappa, h_i^a)
\end{equation}

The attention mechanism computes the attention weight according to the similarity between the query $h_i^l$ and each key $\kappa_i^a$, and weighted sum of the values is then obtained using the attention weight.

\begin{equation}
h_i^{\widetilde{a}} = Attention(h_i^l, \kappa^a, h^a)    
\end{equation}

The resulting aligned acoustic hidden vector is concatenated with lexical encoder output and given as input to the softmax function as explained in Section \ref{sec:ssl}.


\section{Experiments}
\label{sec:exp_setup}
\subsection{Data}

We conduct our experiments on English Fisher corpus \cite{cieri2004fisher}. The training data consists of 348 hours of conversational telephone speech where as the dev and test sets each consists of around 42 hours. To prepare the data splits, we took a subset of the full Fisher corpus to only include segments of a minimum length of six words in our data sets. Punctuation classes in the Fisher corpus are highly unbalanced (see Table 1), which is typical for conversational speech. Fisher corpus has separate time-annotated and punctuated transcripts. For the forced alignment fusion experiments, we need to compute the word boundary information from time-annotated transcripts using a pretrained acoustic model. For this purpose, we trained a TDNN-LSTM acoustic model with lattice-free Maximum Mutual Information (MMI) criterion [27] using the Kaldi ASR toolkit \cite{povey2011kaldi} and the same model is used for obtaining ASR transcriptions on test data. We restored punctuation marks like periods, commas, and question marks to the time-annotated transcripts by aligning them with the corresponding punctuated transcript. 

\begin{table}[h]
\caption{Distribution of punctuation classes in Fisher
corpus.}
\vspace{-2mm}
\centering
\begin{tabular}{ccc}
\hline
Class & Count & Percentage \\
\hline
No punctuation & 2,962,489 & 80.58 \\
Comma (,) & 70,927 & 11.83 \\ 
FullStop (.) & 362,166 & 6.26 \\
Question mark (?) & 56,128 & 1.33 \\
\hline
\end{tabular}
\vspace{-4mm}
\label{tab:classpunct}
\end{table}


\subsection{Acoustic features}

In addition to pretrained wav2vec features, we also experimented with two other prosodic features: pitch and melspec. The prosodic features are computed using a 25ms frame window with 10ms frame shift. We extracted F0 features based on Kaldi pitch tracker method \cite{ghahremani2014pitch}, a highly modified version of the getf0 (RAPT) algorithm using Kaldi ASR toolkit \cite{povey2011kaldi}. Each frame is represented by 4-dimensional features consisting of - probability of voicing (pov) i.e the warped Normalized Cross Correlation Function(NCFF), normalized log pitch (the log-pitch with pov-weighted mean subtraction over 1.5 second window), delta pitch (time derivative of log-pitch) and raw log pitch. We also use a 80-dimension mel-scale spectrograms as alternative to pitch features as they have shown to transfer prosody well in text-to-speech systems \cite{skerry2018towards}. 

For unsupervised wav2vec feature extraction, we train a wav2vec-large model\footnote{https://github.com/pytorch/fairseq/blob/master/examples/wav2vec} \cite{schneider2019wav2vec} on a 348-hour fisher audio corpus. For training, we preprocess the audio files by splitting each file into separate files of 10 to 30 seconds in length. The model is trained with the objective of constrastive loss and had 12 convolutional layers with skip connections. The output is a 512-dimensional unsupervised wav2vec representation.

\begin{table}[t]
\caption{F1 scores for punctuation prediction using various acoustic features and two different fusion techniques; NP: No punctuation; FS: Fullstop; QM: Question mark;}
\vspace{-2mm}
\begin{tabular}{ccccccc}
\hline
Model & Fusion & Feat & NP & Comma & FS & QM \\
\hline
BLSTM   & - & - & 96.2  & 69.4 & 66.1 & 74.0  \\
BERT & - & - & 96.5 & 71.3  & 71.1 & 78.4 \\
\hline
MuSe & FA & pitch & 97.3 & 74.1  & 74.6 & 80.4  \\
     &    & melspec & 97.4 & 74.2  & 74.6 & 80.5  \\
     &    & wav2vec & 97.5 & 75.6  & 75.6 & 81.3  \\
\hline
MuSe & Att & pitch & 97.3 & 73.5  & 73.4 & 79.0  \\
     &     & melspec & 97.4 & 73.5  & 73.4 & 80.1  \\
     &     & wav2vec & 97.5 & 75.5  & 73.4 & 81.3  \\
\hline
\end{tabular}
\vspace{-4mm}
\label{tab:baseline}
\end{table}

\subsection{Model Configurations}
Our primary baseline model is a 4-layer BLSTM based on the work of Zelasko et. al. \cite{zelasko2018punctuation} with each layer having 128 weights in each direction. We also train another lexical only model which is a pretrained truncated BERT model \cite{devlin2018bert} consisting of 6 transformer self attention layers with each hidden layer of size 768. The proposed (\textit{MuSe}) model consists of a lexical encoder which is a pretrained truncated BERT model. The acoustic encoder used for learning task specific embeddings consists a convolutional layer of kernel size 5 and an LSTM hidden layer of size 256. We use a learning rate of 0.00002 and a dropout of 0.1 for the truncated BERT and MuSe models. For all experiments with pretrained lexical encoder, we use a subword vocabulary size of 28k\footnote{Although lexical encoder could be further pretrained on Fisher data, we didn't investigate it in this paper.}.

\section{Results}
\label{sec:results}

\subsection{Results using Multimodal framework}

First, we compare the performance of our proposed multimodal architecture (\textit{MuSe}) with baseline BLSTM model and lexical only BERT model (see Table \ref{tab:baseline}). As expected, pure lexical BERT model outperformed BLSTM in all punctuation marks. We notice significant improvements (5\%, and 4\%) in Fullstop and Question Mark under F1 metric. This indicates that finetuning a pretrained lexical encoder for punctuation task outperforms the recurrent models that are trained from scratch and is synonymous with several other downstream tasks that are finetuned with BERT \cite{devlin2018bert, chen2019bert, yang2019end}. 

We now compare lexical only models to multimodal fusion models trained on three different features: pitch, melspec and wav2vec. Overall, we observe that using any kind of acoustic information helped in improving punctuation prediction across all three classes (Fullstop, Comma and Question Mark). This denotes that the fusion of acoustic features is still beneficial in conjunction with state-of-the-art pretrained lexical encoders as they model different aspects of punctuation. 

Among acoustic features, pitch and melspec have shown similar performance improvements, except in Question mark when attention is used for fusion. This is understandable given that both pitch and melspec feature are extracted from audio using signal processing techniques and have been used as prosodic futures in the past \cite{skerry2018towards}. Unsupervised wav2vec features proved to be the best among all acoustic features for multimodal fusion and its performance is significantly better (p$<$0.01) than semi-supervised lexical only BERT model with an absolute improvement of 4\% on Comma, 2\% on Fullstop and 3\% on Question Mark. Comparing fusion techniques, we observe that forced alignment (\textit{FA}) fusion performs slightly better than attention (\textit{Att}) based fusion for Fullstop while the performance is similar on Comma and Question Mark. We hypothesize that this is because providing explicit acoustic information through duration labels helps better prediction of full stop as opposed to implicit learning through attention mechanism. Although our results are not directly comparable with the results provided in \cite{zelasko2018punctuation} on Fisher corpus (as the splits are different), we achieved better performance in all classes of punctuation. 


\subsection{Streaming models}

We have conducted experiments to study the real time (streaming) performance of the proposed model when there is no future context. For the purpose of this experiment, we perform upper triangle masking (similar to transformer \cite{vaswani2017attention} decoder layers) in self attention layers of lexical encoder to mask the right side context. Since the acoustic encoder is unidirectional, we did not make any further changes. For the experimental results presented in Table \ref{tab:streaming}, we have used forced alignment as fusion technique and wav2vec as acoustic features. We also trained an additional 4 layer LSTM model for baseline comparison. Similar to bidirectional models, the pretrained BERT model performs $\sim$2\% better than LSTM model thus proving that pretraining also helps in learning better representations in the absence of right side context. We also observe that adding acoustic information leads to an additional $\sim$3\% improvement over the lexical BERT model, confirming the effectiveness of our proposed approach for streaming usecase. 

\begin{table}[h]
\caption{F1 scores for punctuation prediction (streaming).}
\vspace{-2mm}
\centering
\begin{tabular}{ccccccc}
\hline
Model & NP & Comma & FS & QM \\
\hline
LSTM   & 95.4  & 67.6 & 68.7 & 72.3  \\
BERT & 95.6 & 68.5  & 69.7 & 75.1 \\
MuSe & 96.3 & 72.1  & 72.3 & 78.6 \\
\hline
\end{tabular}
\vspace{-4mm}
\label{tab:streaming}
\end{table}
\subsection{Ablation study: data sizes}

We perform experiments on varying data sizes to study the effectiveness of our proposed approach. We experimented with data sizes of 1 hour, 10 hours and 100 hours and the results are reported in Table \ref{tab:data_sizes}. For this study, we compared lexical only models (BLSTM and BERT) with our best performing model (MuSe) from Table \ref{tab:baseline}. As expected, the performance of all three approaches were improved with the increase in amount of data size for both reference and ASR outputs. However, our proposed model (\textit{MuSe}) fared significantly better than other two lexical models when trained on smaller datasets (1 hour and 10 hour). For comparison, our model achieved  $\sim$4-18\% absolute improvement in F1 score when trained on 1 hour of audio and text corpus. 

Both pretrained models (lexical BERT and \textit{MuSe})  performed very well on reference transcripts when compared with BLSTM models. However, the gap was significantly reduced when tested on ASR outputs. Our proposed model (\textit{MuSe}) performed better than lexical BERT due to the fusion of acoustic features but the performance gap on ASR outputs is still quite evident. This shows that although pretrained models perform well on reference transcripts with smaller datasets, they are not yet robust to ASR errors. This is due to the fact that pretrained masked language models like BERT were trained only on reference transcripts and have not seen the grammatical errors that are introduced by ASR. 


\begin{table}[t]
\caption{F1 scores for punctuation on varying data sizes.}
\vspace{-2mm}
\begin{tabular}{lcccccccc}
\hline
\multicolumn{1}{l}{} & \multicolumn{1}{c}{} & \multicolumn{2}{c}{Comma} & \multicolumn{2}{c}{FS} & \multicolumn{2}{c}{QM} \\
Model & hours & Ref & ASR & Ref & ASR & Ref & ASR \\
\hline
BLSTM & 1 & 49.7 & 49.8  & 43.9 & 44.3 & 29.7 & 28.0 \\
BERT &  & 54.5 & 54.0  & 51.2 & 50.8 & 42.5 & 35.7 \\
MuSe &  & 58.8 & 57.4  & 55.2 & 55.2 & 48.3 & 40.1 \\
\hline
BLSTM & 10 & 60.9 & 58.1  & 53.5 & 51.9 & 46.3 & 42.4 \\
BERT &  & 65.6 & 62.0  & 62.5 & 58.9 & 61.6 & 57.8 \\
MuSe &  & 68.2 & 63.5  & 65.9 & 62.9 & 72.9 & 60.8 \\
\hline
BLSTM & 100 & 68.9 & 62.8  & 66.1 & 63.4 & 72.8 & 65.6 \\
BERT &  & 70.1 & 65.5  & 69.3 & 64.8 & 75.8 & 68.6 \\
MuSe &  & 71.6 & 66.2  & 70.8 & 65.7 & 77.1 & 69.5 \\
\hline
\end{tabular}
\label{tab:data_sizes}
\vspace{-4mm}
\end{table}

\subsection{Robustness to ASR errors}
\label{ssec:robustness}

We have seen that models trained on reference transcripts did not perform that well when tested on ASR outputs. To make models more robust against ASR errors, we perform data augmentation with ASR outputs for training \cite{sunkara2020robust}. For punctuation restoration, we use edit distance measure to align ASR hypothesis with reference punctuated text and restore the punctuation from each word in reference transcription to hypothesis. If there are words that are punctuated in reference but got deleted in ASR hypothesis, we restore the punctuation to previous word. We performed experiments with data augmentation using N-best lists and the results are reported in Table \ref{tab:robustness}.

From the results, it is evident that the models trained purely on reference transcripts were outperformed by models trained on augmented text (both reference and ASR outputs). The last three rows from Table \ref{tab:robustness} indicate that the data augmentation approach yielded better performance in all classes of punctuation. Overall, data augmentation with 3-best lists gave the best performance. Question mark improved by 6\% in F1 score by performing data augmentation. The improvement might be due to increased number of training examples in augmented data. 

\begin{table}[h]
\vspace{-2mm}
\caption{Comparison of F1 scores for punctuation with models trained on reference transcripts and ASR augmented data.}
\vspace{-3mm}
\begin{tabular}{ccccccc}
\hline
Model & n-best & NP & Comma & FS & QM \\
\hline
BLSTM-Ref & - & 94.5 & 63.5 & 63.8 & 66.7 \\
BERT-Ref   & - & 95.2  & 65.5 & 64.1 & 68.3  \\
MuSe-Ref & - & 95.6 & 67.2  & 66.7 & 70.6 \\
MuSe-ASR & 1-best & 95.8 & 68.5  & 69.0 & 75.7 \\
& 3-best & 95.6 & 69.0  & 69.5 & 76.4 \\
& 5-best & 95.5 & 67.3  & 66 & 76.3 \\
\hline
\end{tabular}
\label{tab:robustness}
\vspace{-6mm}
\end{table}
\section{Conclusions}
\label{sec:conclusions}

We introduced a novel multimodal semi-supervised learning framework which leverages large amounts of unlabelled audio and text data for punctuation prediction. We proposed an alternative attention based multimodal fusion mechanism which is effective, in the absence of forced alignment word durations. Through our data sizes ablation study, we showed how our proposed model is superior in performance to lexical only models on reference transcripts. In order to address the performance gaps on ASR outputs, we presented a robust model that is less affected by ASR errors by performing data augmentation with N-best lists.

\bibliographystyle{IEEEtran}

\bibliography{references}
\end{document}